\documentclass[twocolumn,english,aps,prd,reprint,floatfix,notitlepage,footinbib,preprintnumbers,superscriptaddress,longbibliography]{revtex4-1}
\pdfoutput=1
\usepackage{lmodern}

\usepackage[T1]{fontenc}
\usepackage[latin9]{inputenc}
\usepackage{geometry}
\usepackage{subfigure,lmodern, amsmath,amssymb, graphicx, pifont, adjustbox, bm, xcolor}
\usepackage{amsfonts}
\usepackage{enumitem}
\usepackage{comment}
\usepackage{mathtools}
\usepackage{float}
\usepackage{slashed}
\usepackage{ragged2e}
\usepackage{array}
\usepackage{bbm}
\usepackage{balance}

\usepackage{nameref}

\usepackage{hhline}

\makeatletter\g@addto@macro\bfseries{\boldmath}\makeatother

\makeatletter\newcommand{\labeltext}[2]{%
  \def\@currentlabel{#1}%
  \label{#2}%
}
\makeatother

\usepackage{stackengine}
\usepackage{esint}
\usepackage[unicode=true,pdfusetitle,
 bookmarks=true,bookmarksnumbered=false,bookmarksopen=false,
 breaklinks=false,pdfborder={0 0 1},backref=false,colorlinks=true]
 {hyperref}
\hypersetup{
 pdfauthor={Francesco Calisto, Clifford Cheung, Grant N. Remmen, Francesco Sciotti, Michele Tarquini},
 citecolor=black,linkcolor=black,urlcolor=black}

\newcommand{\appendixref}[1]{\hyperref[#1]{appendix~\ref{#1}}}
\def\equationautorefname~#1\null{eq.\,(#1)\null}
\usepackage{breakurl}
\usepackage[hang,flushmargin]{footmisc} 
\allowdisplaybreaks
\makeatletter

\usepackage{etoolbox}
\apptocmd{\thebibliography}{\justifying\setlength{\leftskip}{7.4mm}}{}{} 
 
 \usepackage{relsize}
\usepackage{babel}

\makeatletter
\def\simgt{\mathrel{\lower2.5pt\vbox{\lineskip=0pt\baselineskip=0pt
           \hbox{$>$}\hbox{$\sim$}}}}
\def\simlt{\mathrel{\lower2.5pt\vbox{\lineskip=0pt\baselineskip=0pt
           \hbox{$<$}\hbox{$\sim$}}}}
\makeatother

\usepackage{changepage}

\newcommand{\be}{\begin{equation}}
\newcommand{\ee}{\end{equation}}
\newcommand{\bea}{\begin{eqnarray}}
\newcommand{\eea}{\end{eqnarray}}
\newcommand{\Fig}[1]{Fig.~\ref{#1}}

\newcommand{\Eq}[1]{Eq.~(\ref{#1})}

\newcommand{\eq}[2]{\be\begin{aligned}#1 \label{#2}\end{aligned}\ee}

\newcommand{\mysec}[1]{\noindent {\bf #1.}---}

\newcolumntype{P}[1]{>{\centering\arraybackslash}p{#1}}

\usepackage{fix-cm}

\definecolor{dartmouthgreen}{rgb}{0.05, 0.5, 0.06}

\begin{document}

\preprint{CALT-TH 2026-014}

\title{Inverse Problem in Effective Field Theory}

\author{Francesco Calisto}
\affiliation{Walter Burke Institute for Theoretical Physics and
Leinweber Forum for Theoretical Physics, California Institute of Technology, Pasadena, CA 91125, USA}
\author{Clifford Cheung}
\affiliation{Walter Burke Institute for Theoretical Physics and
Leinweber Forum for Theoretical Physics, California Institute of Technology, Pasadena, CA 91125, USA}
\author{Grant N.~Remmen}
\affiliation{\scalebox{1}{Center for Cosmology and Particle Physics, Department of Physics, New York University, New York, NY 10003, USA}}    
\author{Francesco Sciotti}
\affiliation{IFAE and BIST, Universitat Aut\`onoma de Barcelona, 08193 Bellaterra, Barcelona, Spain}
\author{Michele Tarquini}
\affiliation{Walter Burke Institute for Theoretical Physics and
Leinweber Forum for Theoretical Physics, California Institute of Technology, Pasadena, CA 91125, USA}

\begin{abstract}

\noindent We show that the tree-level spectrum of heavy particles can be directly extracted from the Wilson coefficients of the corresponding effective field theory at low energies.   This procedure is exact when the number of resonances is finite, and otherwise approximate.  Our results are derived from a new class of analytic dispersion relations that depend nonlinearly on the scattering amplitude and apply to an exceedingly broad class of theories and kinematics.

\end{abstract}

\maketitle

\mysec{Introduction}The space of effective field theories (EFTs) is infinite. This vast freedom is parameterized by an unending sequence of Wilson coefficients encoding the strengths of all possible low-energy interactions. At the same time, our actual world is dictated by some fixed set of fundamental laws, so physics at the very shortest distances must be specified by a single set of equations. Given knowledge of this complete description, a mechanical but laborious calculation then fixes the precise values of the Wilson coefficients in the low-energy EFT.  A natural question then arises: can this mapping be inverted?  We dub this challenge the {\it EFT inverse problem}.

In this paper we show how the ultraviolet spectrum of particles is subtly encrypted within the Wilson coefficients of the EFT.  We then present a simple algorithm for reconstructing the full dynamics of a tree-level ultraviolet completion---spectrum, scattering amplitudes, and all---directly from those Wilson coefficients.

To be very concrete, let us consider an arbitrary tree-level quantum field theory (QFT) composed of a finite number of particles.  The corresponding amplitudes are rational functions of kinematic invariants whose simple poles describe the full spectrum of the theory into the ultraviolet.  The low-energy expansion of these functions define the Wilson coefficients of the corresponding EFT.
Since any rational function is fixed by a finite number of evaluations, a finite number of Wilson coefficients in the EFT should encode the very same data.
Here we derive a universal algorithm that achieves precisely this feat, distilling the ultraviolet structure directly from infrared:
\begin{itemize}

\item[{\it i})] Expand the log derivative of the EFT amplitude,
\eq{
 \frac{d}{ds} \log A(s) = \sum_{k}^\infty c_k s^k,
 }{}

\item[{\it ii})] Build a matrix from these EFT coefficients,
\eq{
\mathbf{c}_r =  \left(
\begin{array}{cccc}
c_{r} & c_{r+1} & c_{r+2} & \cdots  \\
c_{r+1} & c_{r+2} & c_{r+3}  &\\
c_{r+2} & c_{r+3}  & c_{r+4} &  \\
\vdots &  &  &\ddots \\
\end{array}
\right). 
}{cr}

\item[{\it iii})] Count the poles and zeros using the rank,
\eq{
\textrm{rank}(\mathbf{c}_r) =d= d_{\rm poles} + d_{\rm zeros}. 
}{rank_condition}

\item[{\it iv})] Determine the locations of these poles and zeros,
\eq{
\det\left({\mathbf{c}^{(d)}_{r+1}} - \lambda \mathbf{c}_{r}^{(d)}\right)  =0.
}{char_eq}

\end{itemize}
Here, $s$  is the squared center-of-mass energy, $d$ counts poles and zeros of $A(s)$, and $\mathbf{c}_r^{(d)}$ is the $d\times d$ principal submatrix of the $\infty\times \infty$ matrix  $\mathbf{c}_r$.

 Steps {\it i}) and {\it ii}) collate the Wilson coefficients of the EFT into the Hankel matrix  $\mathbf{c}_r$.  While this object is formally infinite, step ${\it iii})$ indicates that it actually has finite rank equal to $d$.  This offers a wonderfully practical litmus test for deducing $d$ explicitly.  In particular, 
simply tabulate more and more Wilson coefficients, building up the $\ell\times \ell$ leading principal submatrix $\mathbf{c}_r^{(\ell)}$ for incrementally increasing values of $\ell$.  At each value of $\ell$, compute the determinant of $\mathbf{c}_r^{(\ell)}$.  Once that determinant vanishes, we know that $\ell=d+1$ and we have determined the number of poles and zeros.
Afterwards, solve the simple characteristic equation in step {\it iv}) for $\lambda$, which as we will show satisfies
\eq{
1/\lambda \;=\; \textrm{locations of poles and zeros},
}{}
thereby dictating the spectrum as well as the entire functional form of the amplitude deep into the ultraviolet. 
Of course, there is a long history of studying the Wilson coefficients of EFTs~\cite{Wan:2024eto,Arkani-Hamed:2020blm,Bellazzini:2020cot} in terms of the mathematics of the moment problem~\cite{schmudgen2017moment}.  The new element here is that our dispersion relation maps the poles and zeros of the amplitude to this moment sequence directly. All dependence on the residues has been sequestered into the locations of the zeros, and the determination of the spectrum has been transformed into finding the solutions of the generalized eigenvalue problem in Eq.~\eqref{char_eq}.

The linchpin of this procedure is a new class of dispersion relations that are {\it nonlinear} in the scattering amplitude.  Our only assumptions are that the amplitude is meromorphic and that its logarithm is polynomially bounded, so our construction is exceedingly insensitive to ultraviolet behavior.   It can thus be applied  where conventional methods fail, for instance in the kinematic regime of {\it hard scattering} or in theories that are {\it nonpolynomially bounded} in energy.
As outlined above, our approach is maximally powerful for QFTs with finite numbers of degrees of freedom but also generalizes to 
 theories with infinite towers of particles, like string theory.

\medskip
\mysec{Dispersion Relation}The central object in our construction is the quotient,
\eq{
Q(s) = \frac{d}{d  s} \log A(s)  =  \frac{A'(s) }{A(s)},
}{}
where $A(s)$ is the four-point tree-level amplitude, suppressing all other kinematic dependences.

Clearly $Q(s)$ exhibits poles when $A'(s)$ is singular or $A(s)$ is vanishing.  At the locus of a pole in $A(s)$ at $s=\mu$, the amplitude and its derivative satisfy
$\lim_{s\rightarrow \mu} A(s)= -R/(s-\mu)+\cdots$
and $\lim_{s\rightarrow \mu} A'(s) = R/(s-\mu)^2 +\cdots$,
and similarly at the locus of a zero at $s=\rho$, $\lim_{s\rightarrow \rho} A(s) = S(s-\rho)+\cdots$ and $\lim_{s\rightarrow \rho} A'(s) = S+\cdots$.
Observe that the residue $R$ and slope $S$ each exactly cancel, so~\footnote{Note that this sum runs over all poles and zeros in the variable $s$ in the amplitude.  Such poles can even arise from long range force carriers in the EFT: for instance, massless exchanges in the $s$ channel, which generate $-1/s$, or massless exchanges in the $t$ channel at fixed $u$, which generate $1/(s+u)$. }
\eq{
Q(s) = -\sum_{\rm poles} \frac{1}{s-\mu}+\sum_{\rm zeros} \frac{1}{s-\rho} +\cdots,
}{Q_poles}
where the ellipses denote terms that are entire in $s$.  Something a bit miraculous has happened~\footnote{See Ref.~\cite{Remmen:2021zmc} for an analogous construction of an amplitude-like object from the logarithmic derivative of the Landau-Riemann $\Xi$ function, encoding the locations of the zeta function's zeros. }: all information except the locations of the poles and zeros has evaporated from singular components of $Q(s)$!

The simple structure of \Eq{Q_poles} motivates us to consider a dispersion relation for $Q(s)$.  Importantly, $Q(s)$ is calculable within the low-energy expansion of the EFT, 
\eq{
Q(s) = \frac{d}{ds} \log A(s) = \sum_{k}^\infty c_k s^k.
}{Q_def}
whose series coefficients $c_k$ will play a crucial role for us.   
While the Wilson coefficients one would measure experimentally are those in the expansion of $A(s)$ rather than $Q(s)$, the coefficients of these expansions are of course bijectively related, with a closed form given in App.~\hyperref[app:AQ]{A}; we will therefore view the $c_k$ themselves as equivalent to Wilson coefficients.
Here the lower limit on $k$ can be negative if there are long range forces. 
We can then relate $c_k$ to the ultraviolet dynamics via a contour integral about the origin of the complex $s$ plane,
\eq{\hspace{-1mm}
c_k&= \frac{1}{2\pi i}\oint\limits_{s=0} \frac{ds}{s^{1+k}} Q(s)\\
 &= {-} \frac{1}{2\pi i}\oint\limits_{s\neq 0}  \frac{ds}{s^{1+k}} \left(  {-}\sum_{\rm poles} \frac{1}{s\,{-}\,\mu}\,{+}\,\sum_{\rm zeros} \frac{1}{s\,{-}\,\rho} \right),
}{}
where $k$ defines the degree of the subtraction.  Evaluating the contour integral on the right-hand side yields a generalized version of Cauchy's argument principle~\footnote{Cauchy's argument principle states that
\begin{equation*}
\qquad\;\;\frac{1}{2\pi i}\protect\oint_\gamma \frac{f'(z)}{f(z)}\,dz = d_{\rm zeros} - d_{\rm poles},
\end{equation*}
where $f(z)$ is meromorphic and $\gamma$ is a closed curve encircling $d_{\rm zeros}$ zeros and $d_{\rm poles}$ poles of $f(z)$.
}
\eq{
c_k&= \sum_{\rm poles} \frac{1}{\mu^{1+k}} - \sum_{\rm zeros} \frac{1}{\rho^{1+k}}  .
}{ck}
Physically, this expression implies our central result that the Wilson coefficients of the EFT directly encode the locations of poles and zeros of the amplitude~\footnote{A higher-degree zero can be accommodated by allowing $\rho$ to run over repeated elements or equivalently by rescaling the corresponding term in \Eq{ck} by the degree of degeneracy.  The same applies for poles, though higher-degree poles are forbidden by locality.}.

In the above equations we have dropped boundary terms at complex infinity for $s$.  This is justified if 
\eq{
\lim_{s\rightarrow \infty} \frac{Q(s)}{s^k} =0,
}{}
for our chosen value of $k$.  We emphasize that this condition is typically {\it exceedingly weak},
since the log of the amplitude in all known cases---including all QFTs and string theories---grows at worst as a polynomial in $s$.
This criterion only fails if the high-energy behavior of the amplitude is {\it doubly exponential} in $s$.

\medskip
\mysec{Field Theory Examples}It will be illuminating to explore the implications of \Eq{ck} in various QFTs.  

To begin, let us consider a toy model describing the exchange of a single particle of mass squared $\mu$ in the $s$ channel.  The corresponding amplitude is
\eq{
A(s) = -\frac{R}{s-\mu}. 
}{}
Since the quotient $Q(s)$ goes to zero at large $s$, we can consider any $k\geq 0$.  From \Eq{ck}, we have
$c_k =1/\mu^{1+k}$,
which confirms the obvious expectation that the Wilson coefficients are infinitely redundant and just reconstruct the geometric series induced by the $s$ channel exchange.

For a theory with two exchanges in the $s$ channel, the story is essentially the same.  The amplitude is
\eq{
A(s) = -\frac{R_1}{s-\mu_1}-\frac{R_2}{s-\mu_2}, 
}{}
and \Eq{ck}  confirms that
\eq{
c_k &= \frac{1}{\mu_1^{1+k}} + \frac{1}{\mu_2^{1+k}} - \frac{1}{\rho^{1+k}} ,
}{}
where the single zero is $\rho =  (R_1 \mu_2 + R_2 \mu_1)/(R_1+R_2)$.

In general, the amplitude for any tree-level, polynomially-bounded QFT takes the form
\eq{
A(s) \propto \prod\limits_{n=1}^{d_{\rm zeros}} (s-\rho_n) \Big/ \prod\limits_{n=1}^{d_{\rm poles}}(s-\mu_n).
}{A_rational}
Here all kinematic invariants other than $s$ have been fixed in such a way that the resulting amplitude is meromorphic in $s$.  This would occur, for instance, at fixed $t$.
By inserting arbitrary values of $\mu_n$ and $\rho_n$ into  \Eq{A_rational}, computing $Q(s)$, and expanding as in \Eq{Q_def}, we precisely confirm Eq.~\eqref{ck}.

\medskip
\mysec{String Theory Examples}Next, let us confirm our result in \Eq{ck} for the case of the Veneziano amplitude for the scattering of open strings~\cite{Veneziano}.  In the context of the superstring, this amplitude is
\eq{
A(s,t) &= -\frac{\Gamma(-s)\Gamma(-t)}{\Gamma(1-s-t)} ,
}{A_Ven}
where for simplicity we have stripped away the gauge invariant prefactor that encodes the external polarization data. It will be instructive to study our results for various choices of kinematics.

\smallskip
\noindent {\it Fixed $t$.} Consider the Regge limit of fixed $t$ and large $s$.  Inserting \Eq{A_Ven} into \Eq{Q_def}, we obtain
\eq{\hspace{-1mm}
Q(s) =&\, \frac{d}{ds} \log A(s,t) \\
=& -\frac{1}{s} + \gamma +\psi(1\,{-}\,t) + \left(\frac{\pi^2}{6} \,{-}\, \psi^{(1)}(1\,{-}\,t) \right)s\\
&+ \left(\zeta(3) \,{+}\, \frac12  \psi^{(2)}(1\,{-}\,t) \right)s^2 +\cdots.
}{Q_Ven}
Since the Veneziano amplitude in Eq.~\eqref{A_Ven} has poles at $s=n$ and zeros at $s=1+n-t$, we find that the terms in Eq.~\eqref{Q_Ven} agree exactly with \Eq{ck}, 
\eq{
c_k^{(t)}(t) &= \sum_{n=1}^\infty \frac{1}{n^{1+k}} -\sum_{n=0}^\infty \frac{1}{(1+n-t)^{1+k}} \\
&= \zeta(1+k) - \zeta(1+k,1-t)
}{eq:ckstring}
after analytic resummation, since the Hurwitz zeta function satisfies $\zeta(1+k,1-t) = (-1)^{k+1} \psi^{(k)}(1-t)/k!$~\footnote{For $k=0$ and $-1$, this sum must of course be handled with care. }.

\smallskip
\noindent {\it Fixed $t/s$.} Conventional dispersion relations fail for {\it hard scattering} in the context of higher-spin towers because the corresponding amplitudes grow as~\cite{Caron-Huot:2016icg}
\eq{
\lim_{s,t\rightarrow+\infty}A(s,t) \sim e^{B(s,t)} ,
}{}
where we have defined
\eq{
B(s,t) = (s\,{+}\,t)\log(s\,{+}\,t) -s\log s - t \log t +\cdots,
}{}
which is familiar from string amplitudes.  This exponential blowup stymies conventional dispersion relations by introducing uncontrolled boundary contributions.

In contrast, our new dispersion relations are perfectly amenable to hard scattering.  Given the fixed ratio,
\eq{
z=t/s,
}{eq:zratio}
we have $B(s,z) =[(1+z)\log(1+z)-z \log z]s+\cdots$, 
 corresponding to a constant quotient,
 \eq{
\lim_{s\rightarrow \infty} Q(s) = (1+z)\log(1+z)-z \log z.
}{} 
Thus, our dispersion relations are applicable for $k\geq 1$.  Expanding the Veneziano amplitude at fixed $z$, we obtain
\eq{
Q(s) =& \frac{d}{ds} \log A(s,zs) \\
=& -\frac{2}{s} -\frac{\pi^2z }{3}s -3\zeta(3) z(1+z) s^2 + \cdots.
}{eq:Qz}
With the choice in Eq.~\eqref{eq:zratio}, the Veneziano amplitude in Eq.~\eqref{A_Ven} has poles at $s=n$ and $n/z$, as well as zeros at $s=(1+n)/(1+z)$, giving us the prediction of \Eq{ck},
\eq{
c_k &= \sum_{n=1}^\infty \frac{1}{n^{1+k}} +\sum_{n=1}^\infty \frac{1}{\left(\frac{n}{z}\right)^{1+k}}-\sum_{n=0}^\infty \frac{1}{\left(\frac{1+n}{1+z}\right)^{1+k}}\\
&= \zeta(1+k) \left(1+z^{1+k} -(1+z)^{1+k} \right).
}{}
Comparing with Eq.~\eqref{eq:Qz}, we find complete agreement.

\smallskip
\noindent {\it Fixed $u$.} The case of fixed $u$ is especially interesting. At fixed $u$, the string amplitude asymptotes to
\eq{
\lim_{s\rightarrow \infty} A(s,u) \sim  s^{u-1}.
}{}
For general fixed $u$, we compute the quotient
\eq{
Q(s)  &= \frac{d}{ds} \log A(s,-s-u) \\
=& -\frac{1}{s} + \gamma +\psi(u) + \left(\frac{\pi^2}{6} + \psi^{(1)}(u) \right)s\\
&+ \left(\zeta(3) + \frac12  \psi^{(2)}(u) \right)s^2 +\cdots ,
}{eq:Qu}
At fixed $u$, the Veneziano amplitude has poles at $s=n$ and $s=-n -u$. While the Veneziano amplitude has ``hidden zeros'' at negative integer $u$ that generalize at higher multiplicity~\cite{Arkani-Hamed:2023swr}, there are otherwise no zeros in $s$ at fixed $u$.
Again computing the prediction of Eq.~\eqref{ck},
\eq{
c_k^{(u)}(u) &= \sum_{n=1}^\infty \frac{1}{n^{1+k}} +\sum_{n=0}^\infty \frac{1}{(-n-u)^{1+k}} \\
&= \zeta(1+k) + (-1)^{k+1} \zeta(1+k,u),
}{cku} 
we find agreement with Eq.~\eqref{eq:Qu}.
Interestingly, for $u=0$, the Wilson coefficients $c_k$ are zero for even $k$ and nonnegative for odd $k$, irrespective of unitarity.

As is well known, the EFT coefficients in the expansion of the Veneziano amplitude in Eq.~\eqref{A_Ven} are multiple zeta values expressed as an exponential sum over simple zeta values~\cite{Schlotterer2012,green2019}. Our dispersion relation for the quotient $Q(s)$ provides an alternative understanding of this behavior.

\medskip
\mysec{Spectroscopy}
The above examples illustrate how the poles and zeros of the amplitude exactly fix the coefficients $c_k$.  Here we will see that the converse is also true: the Wilson coefficients can be manipulated to extract the spectrum of poles and zeros.

Consider a general QFT amplitude, which is polynomially bounded and exhibits a finite number of poles and zeros. Our results imply that the $c_k$ are given as moment sums over the poles and zeros of $A(s)$ as in Eq.~\eqref{ck}.
It will be convenient to repackage this sum as
\eq{
c_k &= \sum_{n=1}^d \sigma_n \lambda_n^{1+k}.
}{eq:ckd}
Here $d=d_{\rm poles}+d_{\rm zeros}$, and $\sigma_n= +1$ or $-1$ when $1/\lambda_n$ is a pole or a zero, respectively. We consider the case where the $\lambda_n$ are all distinct, though this can be generalized~\footnote{In the case of higher-degree zeros, we keep the $\lambda_n$ distinct but absorb the degree of degeneracy into the magnitude of $\sigma_n$. 
Meanwhile, if a pole and a zero happen to coincide, then their contributions cancel in $c_k$, the associated $\sigma_n$ effectively vanishes, and the corresponding $\lambda_n$ can be excluded from the sum.
In such cases, the Vandermonde matrix $[\boldsymbol{\lambda}]_{in} = \lambda_n^i$ remains left-invertible.}. The Hankel matrix in \Eq{cr} becomes
\eq{
[\mathbf{c}_{r}]_{ij} = c_{r+i+j} = \sum_{n=1}^d \sigma_n \lambda_n^{1+r+i+j}.
}{}
Formally, the $i$ and $j$ indices extend infinitely, since there is an infinite sequence of Wilson coefficients in the EFT.
By inspection, we see that $\mathbf{c}_{r}$ factorizes into
\eq{
\mathbf{c}_{r} = \boldsymbol{\lambda}\cdot \mathbf{d}_{r}\cdot \boldsymbol{\lambda}^T,
}{eq:ctransform}
where we have defined the $\infty \times d$ Vandermonde matrix $[\boldsymbol{\lambda}]_{in} = \lambda_n^i$ and the $d\times d$ diagonal matrix 
\be 
[\mathbf{d}_{r}]_{nm} = \sigma_n \lambda_n^{1+r}  \delta_{nm}.\label{eq:dmat}
\ee
Since the $\lambda_n$ are by definition distinct, $\boldsymbol{\lambda}$ is left-invertible. The Vandermonde decomposition in Eq.~\eqref{eq:ctransform} preserves rank, so $\mathbf{c}_{r}$ has rank $d$, confirming \Eq{rank_condition}.  

A corollary of this logic is that the $\ell\times\ell$ leading principal submatrix satisfies
$\textrm{rank}(\boldsymbol{c}_r^{(\ell)}) = \min(\ell,d)$
which reduces to \Eq{rank_condition} when $\ell$ is infinite.  This observation implies that the determinant of the matrix is zero beyond a critical value,
\eq{
\det(\boldsymbol{c}_r^{(\ell)}) = 0 \quad \textrm{for} \quad \ell > d.
}{}
Then $d$ is determined by populating the matrix $\boldsymbol{c}_r^{(\ell)}$ until $\ell$ is sufficiently large that the determinant vanishes.

To enumerate these quantities, we solve the characteristic equation in \Eq{char_eq} for $\lambda$.  The argument of the determinant in \Eq{char_eq} is the $d\times d$ submatrix of $\mathbf{c}_{r+1}-\lambda \mathbf{c}_{r} = \boldsymbol{\lambda}\cdot (\mathbf{d}_{r+1}-\lambda \mathbf{d}_{r})\cdot \boldsymbol{\lambda}^T$.
Rewriting the diagonal matrix in the center as $[\mathbf{d}_{r+1}- \lambda \mathbf{d}_{r}]_{nm} = \sigma_n \lambda_n^{1+r} (\lambda_n -\lambda) \delta_{nm}$,
we see that zeros occur precisely when $\lambda = \lambda_n$.  
The solutions of \Eq{char_eq} thus define the entire sequence $\lambda_n$. Inverting the Vandermonde matrix, we have $\mathbf{d}_{r}= \boldsymbol{\lambda}^{-1}\cdot \mathbf{c}_{r} \cdot (\boldsymbol{\lambda}^T)^{-1}$. For odd $r$, we see from Eq.~\eqref{eq:dmat} that the eigenvalues of this matrix precisely dictate $\sigma_n$, which flags whether $1/\lambda_n$ is a pole or a zero. 
We thus find that
\eq{
\sigma_n = {\rm sign}([\mathbf{d}_{r}]_{nn})  \textrm{ for odd } r.
}{eq:sigmasign}
Since we can always choose odd $r$,
this final datum allows us to determine the exact positions of all poles and zeros.

\begin{figure}[t]
\includegraphics[width=\columnwidth]{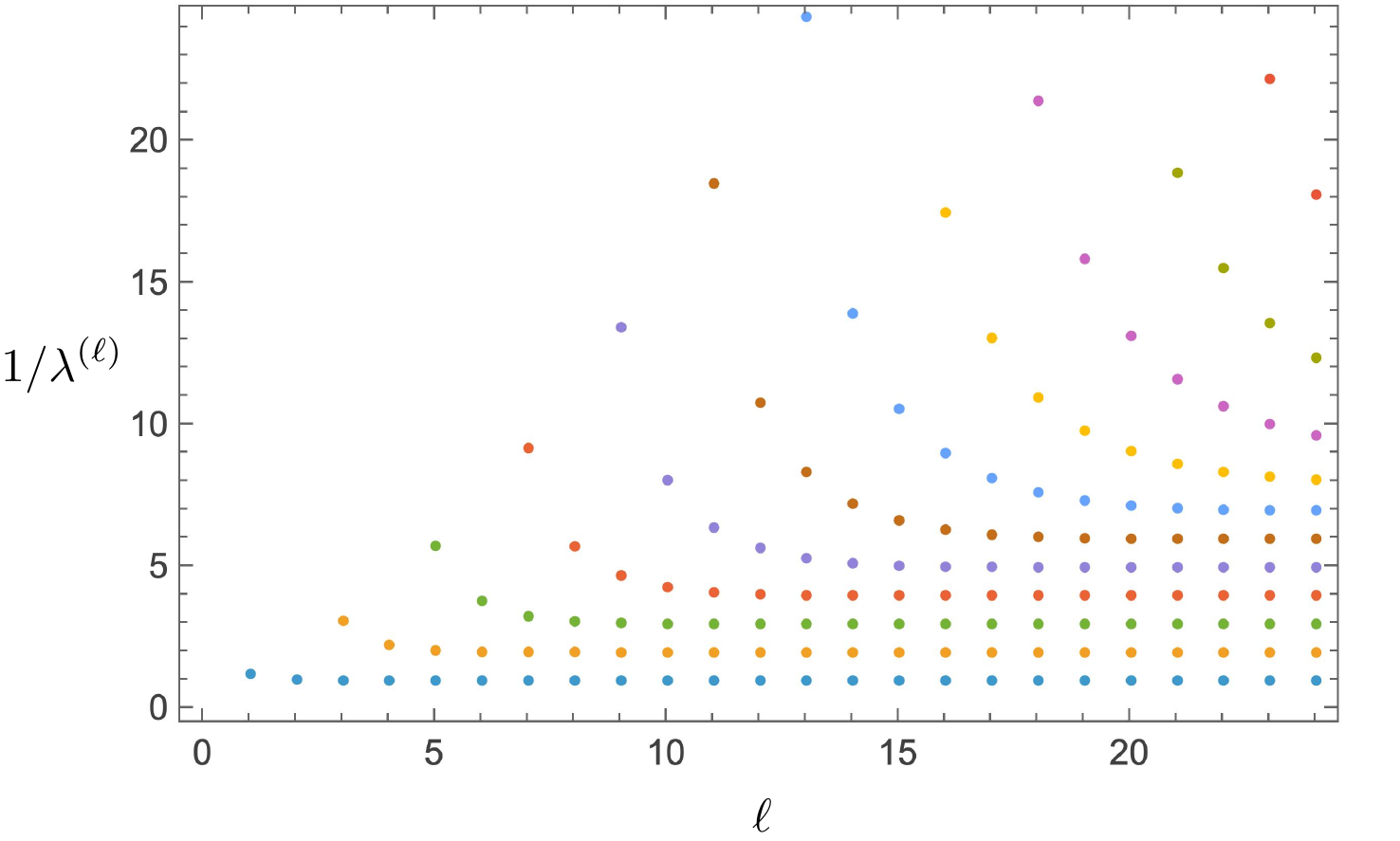}
\caption{The spectrum of string theory is extracted directly from its low-energy EFT.  Given the Wilson coefficients $c_k$ in  \Eq{cku} for $u=0$, construct the $\ell \times \ell$ principal submatrix $\boldsymbol{c}_r^{(\ell)}$, whose entries are $[\mathbf{c}_{r}^{(\ell)}]_{ij} = c_{r+i+j}$.  Solving $\det({\boldsymbol{c}}_{r+1}^{(\ell)} -  \lambda^{(\ell)} {\boldsymbol{c}}_r^{(\ell)}) = 0$ for $r=1$, then plot $1/ \lambda^{(\ell)}$ as a function of $\ell$.   The spectrum approaches that of the string for large $\ell$.}
\label{fig:strings}
\end{figure}

In summary, to extract the complete spectrum of poles and zeros from the EFT coefficients,  solve for $\lambda$ in \Eq{char_eq} to extract all $\lambda_n$.  Then insert $\lambda_n$ into the definition of $\boldsymbol{\lambda}$ to define the matrix ${\bf d}_r$, which determines the sign of $\sigma_n$ from Eq.~\eqref{eq:sigmasign} and specifies $1/\lambda_n$ is a pole or a zero.  While this procedure takes as input the Wilson coefficient expansion of $Q(s)$, the very same algorithm applies to the coefficients of $A(s)$, in which case $\lambda_n$ and $\sigma_n$ in \Eq{eq:dmat} are the mass and residue at each pole.

The above procedure works beautifully if we have access to a sufficient number of Wilson coefficients.  Since the Hankel matrix $\boldsymbol{c}_{r}$ has rank $d$, we need all Wilson coefficients in the leading $d \times d$ principal submatrix $\boldsymbol{c}_{r}^{(d)}$.  But what if we have insufficient data?  Concretely, consider the case where we are privy only to the Wilson coefficients in the leading $\ell\times \ell$ principal submatrix $\boldsymbol{c}_{r}^{(\ell)}$ for $\ell < d$.  Notably, this case subsumes that of string theory, whose infinite spectrum implies $d=\infty$.
The naturally truncated version of \Eq{char_eq} is then~\footnote{A similar mathematical construction was used in Ref.~\cite{Bellazzini:2020cot} to extract the spectrum, albeit in the context of scattering in the forward limit.  In the present work, we directly extract both the poles and zeros of the amplitude at general kinematics.}
\eq{
\det\left({\boldsymbol{c}}_{r+1}^{(\ell)} -  \lambda^{(\ell)} {\boldsymbol{c}}_r^{(\ell)}\right) = 0 \quad \textrm{for} \quad \ell <d.
}{char_eq_trunc}  
As shown in App.~\hyperref[pade]{B}, while $1/\lambda$ labels the poles and zeros of $Q(s)$, the quantity $1/\lambda^{(\ell)}$ labels the poles and zeros of the best-fit rational polynomial for $Q(s)$, known as a Pad\'e approximant.  Even for an infinite spectrum, the convergence properties of Pad\'e approximants~\cite{Nuttall} imply the convergence of $1/ \lambda^{(\ell)}$ to the exact locations of the poles and zeros as $\ell\rightarrow\infty$. 
See \Fig{fig:strings} for the approximate string spectrum extracted from the Wilson coefficients of the Veneziano amplitude in \Eq{A_Ven} for $t=-s$.

\medskip
\mysec{Double Copy}Conventional dispersion relations are linear in the amplitude, so they act simply on {\it sums} of amplitudes.  In contrast, our new dispersion relations are logarithmic, so they play well with {\it products}.  A context where such products arise is the double copy~\cite{Bern:2019prr}.  In particular, the KLT relations~\cite{KLT} recast the Virasoro-Shapiro amplitude~\cite{Virasoro,Shapiro} of closed strings as
\eq{
A^{(\rm VS)}(s,t)&= -\frac{\Gamma(-s)\Gamma(-t)\Gamma(-u)}{\Gamma(1+s)\Gamma(1+t)\Gamma(1+u)} \\
&=\frac{\sin \pi s}{\pi}\,A^{(\rm V)}(s,t)\,A^{(\rm V)}(s,u),
}{klt}
where $A^{(\rm V)}(s,t)$ is the Veneziano amplitude defined in \Eq{A_Ven}. 
Inserting \Eq{klt} into  our dispersion relation, we discover a subtle link between the Wilson coefficients of the open and closed string,
\eq{
c_k^{(\mathrm{VS})}(t)=c_k^{(t)}(t)+c_k^{(u)}(t)+c_k^{(\mathrm{KLT})}.
}{ckKLT}
where $c_k^{(t)}$ and $c_k^{(u)}$ are defined in Eqs.~\eqref{eq:ckstring} and \eqref{cku}. 
Here $c_k^{(\mathrm{KLT})}$ are new terms that arise in the dispersion relation because of zeros in the KLT kernel and can be given in terms of the Bernoulli numbers,
\eq{
\frac{d}{ds} \log \sin \pi s
& =  
\frac{1}{s}
+
\sum_{k\,\mathrm{odd}}
\frac{(-1)^{\frac{k+1}{2}}(2\pi)^{k+1}}{(k+1)!}\,B_{k+1}\,s^k \\ &
=
\frac{1}{s}
+
\sum_{k=0}^{\infty} c_k^{(\mathrm{KLT})}\,s^k.
}{}
We have verified \Eq{ckKLT} using the Wilson coefficients for the Veneziano and Virasoro-Shapiro amplitudes.

\medskip
\mysec{Discussion}This work offers numerous avenues for future inquiry.
A natural question regards  the inclusion of spin.  For simplicity we have  focused entirely on scattering processes with external scalars only, but extending to general spin should be straightforward.  In particular, one can strip off all helicity weights at will and apply our methodology to the multiplicative remainders.  For example, given a four-dimensional scattering amplitude of fermions, $A(s,t) = \langle 12\rangle [34] f(s,t)$, we would derive dispersion relations for $f(s,t)$.   While this approach implicitly drops helicity-dependent zeros, another alternative would be to instead manipulate the analytic continuation of the {\it squared amplitude}, $|A(s,t)|^2 = s^2 f(s,t)^2$, which carries no helicity weight whatsoever.

Our results should generalize to tree-level amplitudes at higher multiplicity.  In particular, \Eq{ck} applies equally well to higher-point amplitudes, provided we define the variable $s$ in the dispersion relation as some higher-point kinematic invariant.  We have verified by explicit calculation that  the analogue of \Eq{ck} holds for five-point amplitudes in planar $\phi^3$ theory and six-point amplitudes in the nonlinear sigma model.

On the other hand, a clear limitation of our current framework is the necessity of the tree-level approximation. Since the mechanics of our dispersion relations depend so crucially on the meromorphic structure of tree amplitudes, the extension to loop level is far from clear.  Typical loop amplitudes exhibit  logarithms and polylogarithms, with rare exceptions such as the rational loop amplitudes of Yang-Mills theory~\cite{Henn_2020}  and self-dual gravity~\cite{Bern_1998}, which preserve meromorphicity. We leave this important question to future work.

\bigskip

\noindent {\it Acknowledgments:} 
F.C., C.C.,~and M.T.~are supported by the Department of Energy (Grant No.~DE-SC0011632), the Walter Burke Institute for Theoretical Physics, and the Leinweber Forum for Theoretical Physics. 
G.N.R. is supported by the James Arthur Postdoctoral Fellowship at New York University. F.S.~is
supported by the research grants 2021-SGR-00649, PID2023-146686NB-C31, and funding from the European Union NextGenerationEU (PRTR-C17.I1). 

\bibliographystyle{utphys-modified}
\bibliography{LogDispersionRelations}

\pagebreak

\appendix

\mysec{Appendix A: From $A(s)$ to $Q(s)$}\label{app:AQ}The series expansion of $Q(s)= d \log A(s)/ds$ encodes that exact same information as the series expansion for $A(s)$.  
However, any experimental probe will reconstruct the low-energy expansion of $A(s)$ rather than $Q(s)$.
For completeness, we now derive a simple formula relating the series expansions of these functions.
In particular, let us define the EFT expansion of the amplitude,
\be 
A(s) = \sum_{j=0}^\infty b_j s^j,
\ee
for convenience considering the case without a massless pole, since the generalization is straightforward. By explicit calculation, we find that the expansion of $Q(s)$ can be written in terms of the expansion of $A(s)$ as
\be
\begin{aligned}
c_0 &= \frac{b_1}{b_0} \\
c_1 &= -\frac{b_1^2 - 2b_0 b_2}{b_0^2} \\
c_2 &= \frac{b_1^3 -3 b_0 b_1 b_2 + 3b_0^2 b_3}{b_0^3}  \\
c_3 &= -\frac{b_1^4 - 4b_0 b_1^2 b_2 + 2b_0^2 b_2^2 + 4 b_0^2 b_1 b_3 -4 b_0^3 b_4}{b_0^4},
\end{aligned} 
\ee
and so on. 
The numerical coefficients appearing in these equations coincide with those of the multivariable Faber partition polynomials.  In fact, one can find an explicit relation between $b_j$  and $c_k$, given by
\be
\hspace{-1mm} c_k = -(k\,{+}\,1) \!\!\!\!\!\!\!\!\!\! \sum_{\substack{m_i \in \mathbb{Z}_{\geq 0}\\ \sum_{i=1}^{k+1} i m_i = k+1}} \!\!\!\!\!\!\!\!\!\!\Gamma({\textstyle\sum_{l=1}^{k+1}m_l}) \prod_{j=1}^{k+1} \frac{(-b_j/b_0)^{m_j}}{m_j!},\hspace{-1mm}
\ee 
where the right-hand side sums over partitions of $k\,{+}\,1$.
Importantly, any $c_k$ only depends on $b_j$ for $j\leq k\,{+}\,1$.  Reconstruction of the spectrum of poles and zeros requires knowledge of some leading set of the $c_k$, which translates into knowledge of some leading set of the $b_j$ and so might in principle be accessed experimentally.  

\bigskip

\mysec{Appendix B: Pad\'e Approximant Spectroscopy}\label{pade}Our construction is closely related to the methodology of Pad\'e approximants~\cite{Nuttall,Sokolovski:2011odo,PEROTTI201895}.
Let us define the paradiagonal Pad\'e approximant of order $[(\ell+r-1)/\ell]$ to the function $Q(s)$ as   $P(s) =N(s)/D(s)$
for polynomials $N(s)$ and $D(s)$ of degree $\ell+r-1$ and $\ell$, respectively.
We then have by definition $D(s)Q(s)-N(s)=O(s^{2\ell+r})$.
For convenience, let us assume that $Q(s)$ is regular at $s=0$; the analogous construction with a massless pole is straightforward.
Writing 
\eq{
\hat Q(s) = \frac{Q(s)-\sum_{k=0}^{r-1}c_{k}s^{k}}{s^r} = \sum_{k=0}^{\infty} c_{k+r}s^k,
}{}
the Pad\'e approximant to $\hat Q(s)$ of order $[(\ell-1)/\ell]$ has the same denominator $D(s)$, since \be 
\hat D(s)Q(s)-\hat D(s)\sum_{k=0}^{r-1}c_{k}s^{k}-s^{r}\hat N(s)=O(s^{2\ell+r})
\ee
implies that $\hat D(s) = D(s)$.
Writing the expansion $\hat D(s) = \sum_{j=0}^{\ell}d_{j}s^{j}$ and matching the coefficient of $s^{\ell},s^{\ell+1},\ldots,s^{2\ell-1}$ in $D(s)\hat Q(s) = \hat N(s)+O(s^{2\ell})$ gives the Pad\'e condition
\eq{
\sum_{j=0}^{\ell}c_{r+i+\ell-j}d_{j}=0,\quad i=0,\ldots,\ell-1.
}{eq:pade_system}
Let $1/y$ be a zero of $\hat D(s)$.
Then $\hat D(s)/(1-ys)$ is a polynomial whose expansion we can write as $\sum_{j=0}^{\ell-1}e_{\ell-1-j}s^{j}$. Equating the two expansions of $\hat D(s)$, we have 
\begin{equation}
\begin{aligned}
d_{0}&=e_{\ell-1},\qquad d_{\ell}= -ye_{0},\\
d_{j}&=e_{\ell-1-j}-ye_{\ell-j},\;\;1\leq j\leq \ell-1.
\end{aligned}\label{eq:dexp}
\end{equation}
Substituting these relations into \Eq{eq:pade_system}, we have $\sum_{j=0}^{\ell-1}\left(c_{r+1+i+j}- yc_{r+i+j}\right)e_{j}=0$,
or in matrix form, 
\be 
[({\bf c}_{r+1}^{(\ell)}-y {\bf c}_{r}^{(\ell)}){\bf e}]_{i}=0,\label{eq:matrixey}
\ee
so ${\bf e}$ is an eigenvector of ${\bf c}_{r+1}^{(\ell)}-y{\bf c}_{r}^{(\ell)}$
with zero eigenvalue, which implies 
\be 
\det({\bf c}_{r+1}^{(\ell)}-y{\bf c}_{r}^{(\ell)})=0.
\ee
Conversely, suppose we have a solution $y$ of Eq.~\eqref{eq:matrixey} for some eigenvector ${\bf e}$.  We can construct a polynomial $\tilde D(s) = (1-ys)\sum_{j=0}^{\ell-1}e_{\ell-1-j} s^j = \sum_{j=0}^\ell \tilde d_j s^j$.
By inspection, we see that the algebraic relations between $\tilde d_j$ and $e_j$ are precisely those satisfied by $d_j$ in Eq.~\eqref{eq:dexp}, so $d_j = \tilde d_j$ and $\tilde D(s) = \hat D(s)$.
Since $1/y$ is a zero of $\tilde D(s)$ by construction, we have $\hat D(1/y)=0$ as well.
Thus, the solutions
$\lambda^{(\ell)}$ of the truncated Hankel matrix problem in Eq.~\eqref{char_eq_trunc} are precisely the inverses
of the poles of the Pad\'e approximant to $Q(s)$ of order $[(\ell+r-1)/\ell]$.

\end{document}